\documentclass[prb,twocolumn,showpacs,preprintnumbers,amsmath,amssymb]{revtex4}

\usepackage{epsfig}
\usepackage{graphicx}
\usepackage{dcolumn}
\usepackage{bm}
\usepackage{wasysym}
\usepackage{marvosym}
\usepackage{times,amsmath,amssymb}

\newcommand{\abs}[1]{{\left|#1\right|}}
\newcommand{\sgn}[1]{{\mathrm {\,sgn}(#1)}}
\newcommand{\tr}{\operatorname{{\mathrm Tr}}}

\def\Xint#1{\mathchoice
  {\XXint\displaystyle\textstyle{#1}}%
  {\XXint\textstyle\scriptstyle{#1}}%
  {\XXint\scriptstyle\scriptscriptstyle{#1}}%
  {\XXint\scriptscriptstyle\scriptscriptstyle{#1}}%
  \!\int}
\def\XXint#1#2#3{{\setbox0=\hbox{$#1{#2#3}{\int}$}
  \vcenter{\hbox{$#2#3$}}\kern-.5\wd0}}

\def\dashint{\Xint-}
\begin{document}

\title{Scanning Tunneling Microscopy currents on locally disordered graphene}
\author{N. M. R. Peres$^1$, Shan-Wen Tsai$^2$, J. E. Santos$^1$, R. M. Ribeiro$^1$}

\affiliation{$^2$ Department of
Physics and Center of Physics, Universidade do Minho, P-4710-057, Braga, Portugal}

\affiliation{$^2$
Department of Physics and Astronomy, University of California, 
Riverside, CA 92521, USA}

\begin{abstract}
We study the local density of states at and around a substituting
impurity and use these results to compute current versus bias
characteristic curves of Scanning Tunneling Microscopy (STM) experiments 
done on the surface of graphene. This allow us to detect the
presence of substituting impurities on graphene. The case of vacancies
is also analyzed. We find that the shape and magnitude of the 
STM characteristic curves depend on the position of the tip and on the
nature of the defect, with the strength of the binging between the
impurity and the carbon atoms playing an important role. Also the
nature of the last atom of the tip has an influence on the shape of the
characteristic curve.
\end{abstract}

\maketitle
 
\section{Introduction}

Graphene\cite{novo1,pnas} consists of a monolayer of carbon atoms forming a two-dimensional honeycomb lattice. It has been intensively studied 
due to its  fascinating physical properties\cite{rmp} and potential applications. 
The honeycomb lattice consists of two triangular sub-lattices and this is responsible for the linear dispersion of the low-energy excitations and for a pseudospin degree of freedom for electrons in graphene. Many of the novel properties of graphene follow from these two facts.
Because of the Dirac spectrum, disorder can have a significant effect on the electronic properties of graphene, the effect being especially strong when the chemical potential crosses the Dirac point. Extrinsic disorder in graphene can be in the form of impurities,\cite{peres1,pereira,loktev,cheianov,bena} topological defects,\cite{voz1,voz2,yazyev1,yazyev2} edges,\cite{peres2,mucciolo} and substrate corrugations.\cite{substrate} In addition, there is also disorder in the form of intrinsic ripples in the structure of graphene.\cite{ripples1,ripples2,yacoby} 
Disorder in graphene occur naturally, but can also be induced if
this is advantageous, to tailor its transport properties. This is the
case for the recently produced material graphane.\cite{novoselov}
Among
the several possibilities, the replacement of a carbon atom
by a different atom can occur.  Atomic substitution in a carbon honeycomb lattice 
is chemically possible for boron (B) and nitrogen (N) atoms. 
There have been several experimental studies of B and N 
substitution in highly-oriented pyrolytic graphite, \cite{mele,endo} 
graphitic structures, \cite{stephan} and nanoribbons. \cite{ze}

In a previous publication\cite{peres2007} the problem of chemical substitution in graphene has been considered, and the local density of states (LDOS) and local electronic structure and charge distribution have been numerically calculated. In this work, we extend the calculation of the spatial dependence of the LDOS at and around the impurity using analytical methods and extending 
the calculations for energies way beyond the Dirac point, an essential ingredient
for the calculation of Scanning Tunneling Microscopy (STM) currents at finite bias.
Using these results we perform theoretical calculations of the tunneling current versus bias characteristic curves of STM studies on graphene surfaces with dilute substitution impurities. The STM technique is one of the most powerful tools for studying surfaces. This is particularly convenient for the study of graphene, which is two-dimensional and exposed to the STM tip. There has been several STM studies of both graphene grown epitaxially on SiC,\cite{stm_rutter,stm_mallet,stm_brar} and mechanically exfoliated graphene on SiO$_2$.\cite{stm_ishigami,substrate,stm_geringer,stm_zhang,stm_deshpande} 
Our results can be used to interprete the STM signal when the tip comes close to a
substituting impurity. A related study, computing the LDOS for point deffects in graphene,
using first principle methods, was recently performed.
\cite{charlier} A study of the LDOS starting from the Dirac equation considering the
effect of magnetic impurities was also recently considered.\cite{Balatsky}
The effect of local potentials with finite strength was studied in Ref. \onlinecite{Loktev}.

In Section \ref{green} we present the tight binding model and the Green's function formalism used. 
For a single localized substitutional impurity, the Green's function can be obtained exactly in closed analytical form. In Section \ref{ldos} we discuss results for the LDOS  and in Section \ref{current} the results for the tunneling current are presented. Section \ref{conclusion} contain further discussions and conclusions.

\section{Calculation of the Green's function for disordered graphene}
\label{green}

\subsection{Basic definitions}
 
\begin{figure}[htf]
\begin{center}
\includegraphics*[width=8cm]{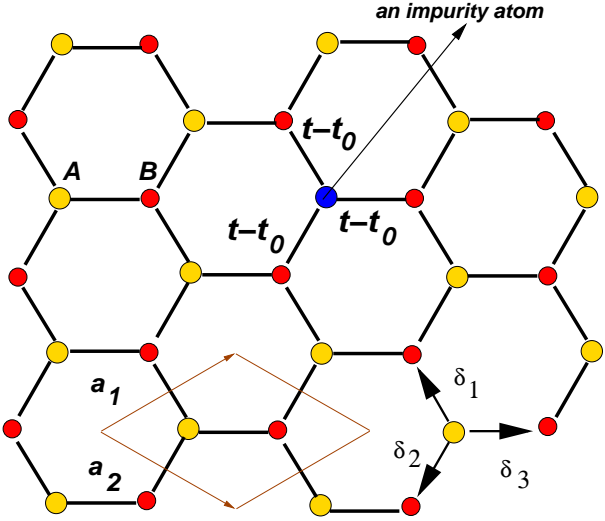}
\end{center}
\caption{(color on line)
\label{Fig_lattice} The honeycomb lattice with a substituting atom replacing 
a carbon atom. The local hopping parameter changes from $t$ to $t-t_0$
around the $A$ atom. We assume all over $t=3$ eV.
\label{fig:lattice}}
\end{figure}

Let us consider that a carbon atom on the otherwise perfect lattice of graphene
has been replaced by a different type of atom. What will happen is a renormalization
of both the on-site energy and the hoping parameter between that atom and the
nearest neighbor carbon atoms. If the hoping parameter of clean graphene is $t$
then the hoping between the impurity atom and the carbon atoms can be parameterized
by an additional constant denoted $t_0$, as shown in Fig. (\ref{fig:lattice}). 
The value of $t_0$ can be varied: if the impurity atom
is strongly coupled to the carbon atom, $t_0$ will be negative; if, on the other hand,
the coupling is weak $t_0$ will be positive. The Hamiltonian of the system will be that
of clean graphene and a term representing the renormalization of the hoping, as
explained below.

Let us first introduce some definitions for latter use.
The honeycomb lattice has a unit cell represented in Fig. \ref{Fig_lattice}
by the vectors $\bm a_1$ and $\bm a_2$, such that 
$\vert \bm a_1 \vert=\vert \bm a_2 \vert=a$, with $a\simeq 2.461$ \AA. 
In this basis any lattice vector $\bm R$
is represented as
\begin{equation}
\bm R= n\bm a_1 + m\bm a_2\,,
\label{Rvec}
\end{equation}
with $n,m$ integers. 
In cartesian coordinates one has,
\begin{equation}
\bm a_1 = \frac {a_0} 2 (3,\sqrt 3,0)\,,
\hspace{.2cm}
\bm a_2 = \frac {a_0} 2 (3,-\sqrt 3,0)\,,
\end{equation}
where $a_0=a/\sqrt 3$ is the  carbon-carbon distance.

If periodic boundary conditions
are used, the Bloch states are characterized by
 momentum vectors of the form
\begin{equation}
\bm k= \frac {m_1}{N_1}\bm b_1 + \frac {m_2}{N_2}\bm b_2\,,
\end{equation}
with $m_1$ and $m_2$ a set of integers running from 
$0$ to $N_1-1$ and from 0 to $N_2-1$, respectively. The numbers
$N_1$ and $N_2$ are the number of unit cells along
the $\bm a_1$ and $\bm a_2$ directions, respectively.
The total number of unit cells is, therefore, $N_c=N_1N_2$.
The reciprocal lattice vectors are given by:
\begin{equation}
\bm b_1= \frac {2\pi}{3a_0} (1,\sqrt 3,0)\,,
\hspace{.2cm}
\bm b_2= \frac {2\pi}{3a_0} (1,-\sqrt 3,0)\,,
\label{recv}
\end{equation}
The  vectors connecting any $A$ atom to its nearest
 neighbors read: 
\begin{eqnarray}
\bm \delta_1 &=& \frac {a_0} 2 (-1,\sqrt 3,0)=\frac 1 3 (\bm a_1-2\bm a_2),\\
\bm \delta_2 &=& \frac {a_0} 2 (-1,-\sqrt 3,0)=\frac 1 3 (\bm a_2-2\bm a_1), \\
\bm \delta_3 &=& a_0 (1,0,0)=\frac 1 3 (\bm a_1+\bm a_2)
\label{nnv}
\end{eqnarray}

Using the above
definitions the Hamiltonian for this problem can be written as
\begin{eqnarray}
H_0 &=& -t\sum_{\bm R}[a(\bm R)b^\dag(\bm R) + a(\bm R)b^\dag(\bm R-\bm a_2)
\nonumber\\
&+&
 a(\bm R)b^\dag(\bm R-\bm a_1)
+ h.c.
]\,,
\end{eqnarray} 
which represents clean graphene (the spin index is omitted for simplicity of writing),
and the perturbation due to the impurity atom is 
\begin{eqnarray}
\label{Vi}
V_i &=& t_0[a(0)b^\dag(0) + a(0)b^\dag(-\bm a_2)\nonumber\\
&+&
 a(0)b^\dag(-\bm a_1)
+ h.c.
]\,.
\end{eqnarray}
It is assumed that the impurity atom is in the unit cell $\bm R=0$, but there
is no loss of generalization due to this choice.
In the particular case of $t_0=t$, the scattering term
$V_i$ represents a vacancy, since the impurity atom
is completely decoupled from the carbon atoms. It is important to keep in mind
that in a given unit cell both $A$ and $B$ type of atoms are described 
by the same vector $\bm R$. 

The calculation of the electronic properties of graphene requires the
calculation of the corresponding Green's functions,
whose definitions are
\begin{eqnarray}
G_{aa}(\bm k,\bm q,\tau) &=&
   -\left< T \left[a_{\bm k}(\tau) \: a^\dag_{\bm q}(0)\right] \right>\,,\\
G_{bb}(\bm k,\bm q,\tau) &=&
   -\left< T \left[b_{\bm k}(\tau) \: b^\dag_{\bm q}(0)\right] \right>\,,\\
G_{ab}(\bm k,\bm q,\tau) &=&
   -\left< T \left[a_{\bm k}(\tau) \: b^\dag_{\bm q}(0)\right] \right>\,,\\
G_{ba}(\bm k,\bm q,\tau) &=&
   -\left< T \left[b_{\bm k}(\tau) \: a^\dag_{\bm q}(0)\right] \right>\,,
\end{eqnarray}
and their Fourier transform to the Matsubara representation are
given in the Appendix \ref{free} for the case of a perfect lattice. Of particular
interest to our calculations is the momentum integral of the retarded diagonal Green's function
\begin{equation}
\bar G_{AA}^0(\omega)=\frac 1{N_c}\sum_{\bm k}G^0_{AA}(\bm k,\omega)\,.
\label{Gbar}
\end{equation}
The integral is best performed using the density of states of the honeycomb lattice.
Since we want to take into account the non-linearity of the bands, which allows us to
describe the properties of the system at large energies and not only close to the
Dirac point, we have to use for the density of states an expression that goes beyond
the usually used linear dependence of this quantity on energy. In a previous work\cite{condgeim08} 
we derived an expansion for the density of states (per unit cell, per spin)
valid for energies up to
$\sim 3$ eV, reading ($E=\hbar\omega$)
\begin{equation}
\rho(E)\simeq \frac {2E}{\sqrt 3 \pi t^2}+\frac {2E^3}{3\sqrt 3 \pi t^4}
+\frac {10E^5}{27\sqrt 3 \pi t^6}\,.
\label{expand_rho}
\end{equation}
The imaginary part of $\bar G_{AA}^0(\omega)$ reads
\begin{equation}
 \Im \bar G_{AA}^0(\omega) = -\frac{\pi}2\rho(\hbar\omega)\,,
\end{equation}
and the real part has the form
\begin{equation}
 \Re \bar G_{AA}^0(\omega)=P_1(\hbar\omega)+P_2(\hbar\omega)\ln\frac{(\hbar\omega)^2}{D_c^2-(\hbar\omega)^2}\,,
\end{equation}
where 
$P_1(x)$ and $P_2(x)$ are polynomial functions given by
\begin{eqnarray}
 P_1(x)&=&-\frac {x}{3t^2}-\frac{5}{27t^4}\left(
\frac x 2 D^2_c + x^3
\right)\,,\\
P_2(x)&=&\frac {x}{D_c^2} + \frac{x^3}{3t^2D^2_c} + \frac{5}{27D^2_ct^4}x^5\,.
\end{eqnarray}
The energy $D_c$ is a cut-off energy chosen as $D^2_c=\sqrt 3\pi t^2$.

\subsection{Exact Green's Functions}

We now want to determine the exact expressions for the Green's functions in the
presence of the substituting atom. This is best accomplished using the equation of motion
method.
The equations of motion for the Green's functions can be readily established,
and read
\begin{widetext}
\begin{eqnarray}
i\omega_n G_{AA}(\omega_n,\bm k,\bm p)&=&\delta_{\bm k,\bm p} + t\phi(\bm k)
G_{BA}(\omega_n,\bm k,\bm p) - \frac {t_0}{N_c}\sum_{\bm k'}\phi(\bm k')
G_{BA}(\omega_n,\bm k',\bm p)\,,\\
i\omega_n G_{BA}(\omega_n,\bm k,\bm p)&=& t\phi^\ast(\bm k)
G_{AA}(\omega_n,\bm k,\bm p)- \frac {t_0}{N_c}\phi^\ast(\bm k)\sum_{\bm k'}
G_{AA}(\omega_n,\bm k',\bm p)\,,\\
i\omega_n G_{AB}(\omega_n,\bm k,\bm p)&=& t\phi(\bm k)
G_{BB}(\omega_n,\bm k,\bm p) - \frac {t_0}{N_c}\sum_{\bm k'}\phi(\bm k')
G_{BB}(\omega_n,\bm k',\bm p)\,,\\
i\omega_n G_{BB}(\omega_n,\bm k,\bm p)&=& \delta_{\bm k,\bm p} + 
t\phi^\ast(\bm k)
G_{AB}(\omega_n,\bm k,\bm p)- \frac {t_0}{N_c}\phi^\ast(\bm k)\sum_{\bm k'}
G_{AB}(\omega_n,\bm k',\bm p)\,.
\end{eqnarray}
\end{widetext}
The complex number $\phi(\bm k)$ is defined as
\begin{eqnarray}
\phi(\bm k) &=& 1 + e^{i\bm k\cdot\bm a_1} + e^{i\bm k\cdot\bm a_2}\nonumber\\
&=&1+e^{i\bm k\cdot(\bm\delta_3-\bm\delta_1)} + e^{i\bm k\cdot
(\bm\delta_3-\bm \delta_2)}
\,,
\end{eqnarray}
which is the form factor of the three $B$ atoms as seen by an atom in $A$.
The above set of equations can be solved exactly. The fact that the
scattering term $V_i$ depends on $\phi(\bm k)$ and that a single impurity 
breaks particle-hole symmetry of the system implies a complex form 
for $T$-matrix. In fact, the general expression for the Green's functions
does not have exactly the same form as in the case of one-band electrons.
The solution of the equations of motion 
is rather lengthy and in the course of the solution we use the two following identities:

\begin{equation}
i\omega_n \sum_{\bm k} G_{AA}(\omega_n,\bm k,\bm p)=
1 + (t-t_0)\sum_{\bm k} \phi(\bm k) G_{BA}(\omega_n,\bm k,\bm p)\,,
\label{r1}
\end{equation}

and 

\begin{eqnarray}
t_0 z &=&
-[(i\omega_n)^2+ z t_0(t-t_0)]\sum_{\bm k} \phi(\bm k)
G_{BA}(\omega_n,\bm k,\bm p)\nonumber\\
&+& i\omega_n t\sum_{\bm k} \vert\phi(\bm k)\vert^2 
G_{AA}(\omega_n,\bm k,\bm p)\,,
\label{r2}
\end{eqnarray}
with 
\begin{equation}
z=\frac 1 {N_c}\sum_{\bm k}  \vert\phi(\bm k)\vert^2 \,.
\end{equation}
The use of the relations (\ref{r1}) and (\ref{r2})
leads to 
\begin{equation}
\sum_{\bm k} G_{AA}(\omega_n,\bm k,\bm p) = \frac {N_1(\bm p,\omega_n)}{D(\omega_n)}\,,
\label{sumgaa}
\end{equation} 
with the following definitions
\begin{equation}
N_1(\bm p,\omega_n) = (t-t_0)G^0_{AA}(\omega_n,\bm p)
+t_0 \bar G^0_{AA}(\omega_n)\,,
\end{equation}
\begin{eqnarray}
D(\omega_n)
= (t-t_0)(1-t_0/t) +i\omega_n(2t_0-t_0^2/t)\bar G^0_{AA}(\omega_n)\,,
\end{eqnarray}
\begin{equation}
\bar G^0_{AA}(\omega_n) = \frac 1 {N_c}\sum_{\bm k} G^0_{AA}(\omega_n,\bm k)\,,
\label{bargaa0}
\end{equation}
and 
\begin{equation}
\tilde G^0_{AA}(\omega_n) = \frac 1 {N_c}
\sum_{\bm k} 
\vert\phi(\bm k)\vert^2 
G^0_{AA}(\omega_n,\bm k)\,.
\label{tildegaa0}
\end{equation}
The result (\ref{sumgaa}) has the appropriate limiting behavior:
when $t_0\rightarrow t$ one obtains $(i\omega_n)^{-1}$, which agrees
with (\ref{r1}); when  $t_0\rightarrow 0$ one obtains 
$G^0_{AA}(\omega_n,\bm p)$, which is the result for the perfect lattice.

Finally, the exact solution for $G_{AA}(\omega_n,\bm k,\bm p)$, 
considering an arbitrary value of $t_0$,
has the following structure
\begin{eqnarray}
&&G_{AA}(\omega_n,\bm k,\bm p)=\delta_{\bm k,\bm p}
G_{AA}^0(\omega_n,\bm k) + G_{AA}^0(\omega_n,\bm k)
\Sigma(\omega_n) \nonumber\\
 &+& g(\omega_n)+
G_{AA}^0(\omega_n,\bm k)T_{AA}(\bm k,\bm p, \omega_n)G_{AA}^0(\omega_n,\bm p)\,,
\label{GAA}
\end{eqnarray}
with 
\begin{equation}
 g(\omega_n) = \frac 1 {N_c}\frac {t^2_0}t\frac {\bar G_{AA}^0(\omega_n)}
{D(\omega_n)}\,,
\end{equation}
\begin{equation}
 \Sigma(\omega_n) = \frac 1 {N_c}\frac {t_0(1-t_0/t)}
{D(\omega_n)}\,,
\end{equation}

and
$T_{AA}(\bm k,\bm p, \omega_n)$ given by
\begin{equation}
T_{AA}(\bm k,\bm p, \omega_n) = \frac 1 {N_c}t_0\frac 
{t(t_0-t)\vert\phi(\bm k)\vert^2-(i\omega_n)^2}
{i\omega_n D(\omega_n)}\,.
\end{equation}
Clearly, the solution (\ref{GAA}) does not have the simple form of the $T-$matrix
as in the case of the
solution of the scattering problem by a local potential in the single energy-band case.
The calculation for  $G_{BB}(\omega_n, \bm k,\bm p)$ proceeds along the same lines
and gives
\begin{eqnarray}
&&G_{BB}(\omega_n,\bm k,\bm p)=\delta_{\bm k,\bm p}
G_{BB}^0(\omega_n,\bm k)+ \nonumber\\
&+&G_{BB}^0(\omega_n,\bm k)T_{BB}(\bm k,\bm p, \omega_n)G_{BB}^0(\omega_n,\bm p)\,,
\label{GBB}
\end{eqnarray}
with $T_{BB}(\bm k,\bm p, \omega_n)$ given by
\begin{equation}
T_{BB}(\bm k,\bm p, \omega_n) = -\frac 1 {N_c}\frac 
{\phi^\ast(\bm k)(2t-t_0)tt_0\phi(\bm p)}
{i\omega_n D(\omega_n)}\,.
\end{equation}
It is interesting to note that is the case of Eq. (\ref{GBB}) the $T-$matrix
has the traditional form. So, as long as one does not sit on top of the impurity
atom, the scattering equations are similar to those of the scattering of  the
electron gas by a local potential.
The exact results  ($\ref{GAA}$) and ($\ref{GBB}$) is what is need to discuss
the behavior of current versus bias in STM experiments when the microscope tip
comes near to a substituting defect.

\section{Local density of states}
\label{ldos}

As we will see later, the properties of the STM current will depend
on the local density of states below the tip of the microscope. We therefore
have to compute this quantity for different circumstances.
Because each unit cell is identified by a single vector
$\bm R$, the  local density of states (per spin)  at the  atoms $A$ and $B$
of the unit cell
localized in the position $\bm R$ is defined as 
\begin{equation}
\rho_x(\bm R,\omega) = -\frac 1 {\pi N_c} 
{\rm Im} G_{xx}(\bm R,\bm R,\omega)\,,
\label{rho_x}
\end{equation}
with $G_{xx}(\bm R,\bm R,\omega)$ $(x=A,B)$ obtained  from
\begin{eqnarray}
G_{xx}(\bm R,\bm R,\omega_n) &=& 
\sum_{\bm k,\bm p}e^{i(\bm k-\bm p)\cdot \bm R}
 G_{xx}(\bm k,\bm p, \omega_n)\,,
\label{realspaceG}
\end{eqnarray}
after the usual analytical 
continuation $\omega_n\rightarrow \omega + i0^+$ of the Matsubara Green's function.

Let us first consider the case $t_0=t$. With this choice the $A$ atom is disconnected
from the rest of the lattice and in this situation the momentum sum over the full 
$G_{AA}(\omega_n,\bm k,\bm p)$ reads
\begin{equation}
 \sum_{\bm k} G_{AA}(\omega_n,\bm k,\bm p)=\frac 1 {i\omega_n}\,,
\end{equation}
which corresponds to an isolated atom. As a consequence the local
density of states is a delta function and no charge transport can take place
through that atom. In the material this situation never happens and therefore
we can interpreted this result as the case where there is a vacancy in the lattice.
Therefore the presence of a vacancy can be detected by its influence on the
electronic density of states of the neighboring atoms (the $B$ atom in the
case of Fig. \ref{fig:lattice}). Using only the first term in Eq. (\ref{expand_rho})
it is possible to derive a relatively simple expression for the local density of
states at the $B$ atom near the vacancy, reading

\begin{equation}  
\label{eq:LDOS1_Dirac}
  \rho_B (0,\omega) =
   \frac{2}{\sqrt 3\pi t } \abs{\frac{\hbar \omega}{t}}
   \left( 1 - \frac{1}{9} \abs{\frac{\hbar \omega}{t}}^2 
  +  \frac{1}{3} \abs{\frac{t}{\hbar \omega}}^2 L(\omega)
   \right) \,.
\end{equation}
with 
\begin{equation}
 L(\omega)=\left[ 1 + \frac{1}{\pi^2} \ln^2 \! \left( \frac{1}{\sqrt{3} \pi}
   \abs{\frac{\hbar \omega}{t}}^2 \right) \right]^{-1}\,.
\end{equation}
It is also possible to determine general analytical expressions for 
$\rho_A (0,\omega)$ and $\rho_B (0,\omega)$
given
an
arbitrary value of $t_0$, but due to the form of $D(i\omega_n)$
which in this case depends on both $t$ and $t_0$, the final result
is somewhat cumbersome. However, if really needed, it is straightforward to use
the full equations for $G_{AA}$ and $G_{BB}$ and the given expressions for
$\bar G_{AA}(\omega)$ to write down analytical expression for the density of states.

\begin{figure}[htf]
\begin{center}
\includegraphics*[width=8cm]{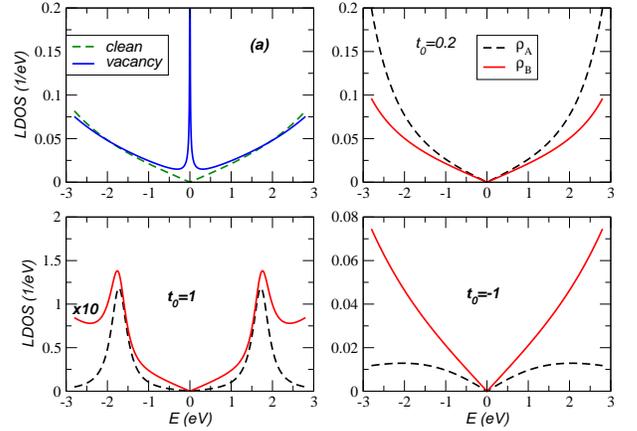}
\end{center}
\caption{(color on line)
Local density of states (LDOS) at the unit cell $\bm R=0$ for the $A$ and $B$ atoms
and different values of $t_0$ in electron-volt. 
In panel (a) we plot the density of states of clean
graphene and the LDOS at the $B$ atom near a vacancy at the $A$ atom of the same
unit cell. For $t_0=1$, $\rho_B(0,\omega)$ is multiplied by 10 for clarity.
\label{fig:LDOS}}
\end{figure}

In Figure \ref{fig:LDOS} we plot the local density of states for different
values of $t_0$. In panel (a) of that figure we represent the density of states
at a $B$ site in the unit cell where a vacancy exists in the $A$ atom of the
same unit cell. The most significant feature is a development of a logarithmic
divergence at zero energy. Therefore the clean density of states 
(also shown for comparison) is strongly
modified by the presence of such a strong potential. For a moderate value of $t_0$
the change of the hoping is not strong and the density of states at both the
$A$ and $B$ atoms retain the linearity of the clean density of states close
to the Dirac point. However, the absolute value of the two density of states
are not the same and a clear deviation from linearity is seen to take place for
lower energies when compared with the clean case. If the impurity atom
binds strongly to the carbon atoms ($t_0=-1$)
there is an increase of the density of states
at the  neighbor atoms at the expenses of the density of states of the impurity.
On the other hand, if the impurity binds weakly to the carbon atoms ($t_0=1$),
the density of states increases at the impurity and strong resonances develop
on the impurity density of states. Clearly these different behaviors of the density of states
will show up in the tunneling current.

It is also instructive to compute the local density of states
as one moves away from the defect. This amounts to perform the Fourier
transforms for finite $\bm R$ in Eq. (\ref{realspaceG}). In the
case of $G_{AA}(\bm R,\bm R, \omega)$ the calculation of the integrals
is facilitated by the fact that the momentum space Green's function
on the sub-lattice $A$
depends on $\vert \phi(\bm k)\vert$ only, whereas in the case of the
Green's function on the $B$ sub-lattice this in not the case.
Besides for finite $\bm R$ the useful relation (\ref{niceproperty})
is valid no more. However, for large $R=\vert \bm R \vert$ values when compared
to $a$, i. e. a way from the defect, the details of the lattice are
no longer important and the behavior of the local density of states at
both the $A$ and $B$ sub-lattices must be similar.  
We will therefore give the density of states as function of $\vert \bm R \vert$
for the $A$ sub-lattice only. Performing the Fourier transforms using the
Dirac cone approximation (this is not a restriction, but helps keeping
the final result not too cumbersome) one obtains (for the retarded function)
\begin{equation}
 G_{xx}(\bm R,\bm R,\omega)=N_c\bar G^0_{AA}(\omega)+
N_c\frac {A^2_c\omega}{2\pi^2v_F^2D(\omega)} \Theta(\omega)\,,
\label{GxxR}
\end{equation}
where $v_F=3a_0t/(2\hbar)$ and $\Theta(\omega)$ can be written as
\begin{eqnarray}
\Theta(\omega)=
 t_0(t_0/t-2)\frac{\omega^2}{2v_F^2}I^2(\omega)\,,
\end{eqnarray}
where 
$I(\omega)$ is defined as
\begin{equation}
I(\omega)=I_0(\omega)-i\pi\sgn{\omega}J_0(\vert\omega\vert R/v_F) 
\end{equation}
with $J_0(x) $  the Bessel function of integer order $n=0$, and finally
$I_0(\omega)$ is the Cauchy principal value of the integral
\begin{equation}
I_0(\omega)=\dashint_0^{k_cR}dx\frac{2xJ_0(x)}{\alpha^2-x^2}\,,
\end{equation}
which can be evaluated using standard numerical methods, and
we also have $k_c=2\sqrt\pi/(\sqrt{3\sqrt{3}}a_0)$,
and
$\alpha = \omega R /v_F$. 
 \begin{figure}[htf]
\begin{center}
\includegraphics*[width=8cm]{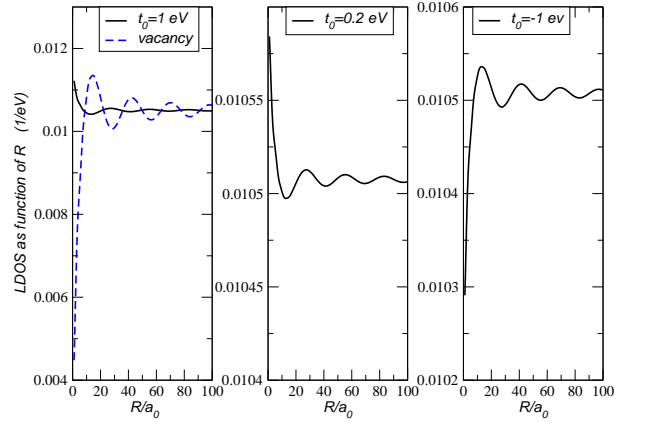}
\end{center}
\caption{(color on line)
Local density of states (LDOS) at distance $R$ from unit cell $\bm R=0$, where the
impurity is located for  different values of $t_0$. 
The energy is $\hbar\omega=0.5$ eV. The case of the vacancy is also illustrated.
\label{fig:LDOS_func_R}}
\end{figure}
In Figure \ref{fig:LDOS_func_R} we depict the local density of states at the carbon
atoms located in the $A$ sub-lattice, computed using Eq. (\ref{rho_x}).
The typical oscillations in the density of states close to impurity centers
are clearly seen. Note that as one moves away from the
impurity the density of states approaches that of the clean system. The Friedel oscillations
in graphene can be obtained from integrating the density of states up to the Fermi energy.

\section{Calculation of the tunneling current}
\label{current}

We now want to compute the tunneling current between the tip of an
STM microscope and graphene, when the tip is closed to an impurity atom.
We model the tip by a one dimensional tight-binding system, a standard
approach.\cite{Mujica,Hong} Our results
will not depend significantly from this choice as long as we assume a large
value for the hoping parameter of the electrons in the tip. This choice
essentially corresponds to represent the tip by a metal with a large bandwidth.
Given the above, the Hamiltonian for the tip has the form
\begin{equation}
 H=-V\sum_{n=-\infty}^{-1}[c^\dag(n-1)c(n)+c(n)c^\dag(n-1)]+ H_0\,,
\label{tip}
\end{equation}
and $H_0$ representing the Hamiltonian of the last atom of the tip.
This choice corresponds to the assumption that the surface atom of the
tip has a different nature from the atoms in the bulk of tip.
We therefore represent $H_0$ by
\begin{equation}
 H_0=\epsilon_0c^\dag(0)c(0)-W_1[c^\dag(-1)c(0)+c(0)c^\dag(-1)]\,.
\end{equation}
It is essential that the Hamiltonian of the tip to be represented
by a semi-infinite metal, since otherwise there would be electrons reflected
at the far end of the tip modifying in this way the tunneling current.
Finally we need to include the tunneling of the electrons of the tip to
graphene. There is a number of ways we can do this. Here we assume that
the coupling is made directly either to the impurity atom or to
the next neighbor carbon atom. This choice corresponds to probing the local
electronic properties at or around the impurity.
More general types of coupling are easily included
in the formalism. We write this coupling as
\begin{equation}
H_T=-W_2[c^\dag(0)d(0)+d^\dag(0)c(0)]\,, 
\end{equation}
where the operator $d(0)$ can represent either the impurity atom
at the $A$ sub-lattice or the carbon atom at the $B$ sub-lattice.

Since the Hamiltonian of the problem is bilinear we can write it
in matrix form (of infinite dimension) as 
\begin{equation}
 H=
\left[
\begin{array}{ccc}
H_b & V_L & 0\\
V^\dag_L & H_0 & V^\dag_R\\
0 & V_R & H_g
\end{array}
\right]
\label{hamiltonian33}
\end{equation}
where the matrices $V_L$ and $V_R$  represent the coupling of the last atom in the tip of
the STM microscope to the bulk of the tip and to graphene, respectively,
and $H_b$ and $H_g$ stand for the bulk Hamiltonians of the tip and
of graphene, including the impurity potential (\ref{Vi}), respectively.

The tunneling is a local property, controlled by the coupling of the last atom of the tip
to the bulk atoms and to graphene. 
Since we want to compute local quantities, this
is best accomplished using Green's functions in real space. The full Green's function of the system
is defined by
\begin{equation}
(\bm 1 E+i0^+-H)G^+=\bm 1\,, 
\end{equation}
where we have chosen the retarded function (denoted with the $+$
superscript), and $\bm 1$ is the identity matrix. The matrix form of the Green's function is  
\begin{equation}
G^+= \left[
\begin{array}{ccc}
 G_{bb}  & G_{b0} & G_{bg}\\ 
G_{0b}  & G_{00} & G_{0g}\\
G_{gb}  & G_{g0} & G_{gg}
\end{array}
\right]\,.
\end{equation}
The quantity of interest is $G_{00}$, which can be shown to have the form 
\begin{equation}
G_{00}^+= (E+i0^+-\epsilon_0-\Sigma_L^+-\Sigma_R^+)\,, 
\end{equation}
where the matrices $\Sigma_L^+$ and $\Sigma_R^+$ are the self energies and have the form
\begin{equation}
 \Sigma_L^+=W_1^2G^+_s\,,
\hspace{0.5cm}
\Sigma_R^+=W_2^2G^+_{xx}\,,
\end{equation}
where the Green's functions $G^+_s$ and $G^+_{xx}$ are the surface Green's function
of the Hamiltonians $H_b$ and $H_g$ at the impurity unit cell
($x=A,B$), respectively.  Note that the quantity $G^+_{xx}$
is computed using Eq. (\ref{GxxR}) making $\bm R=0$.
It is possible to find a close form for $G^+_{xx}$, as we have
shown in the previous section. Also for $G^+_s$ a close form exists 
\cite{Dy,Sankey,peresdot}
\begin{equation}
\label{Gbb}
G_s^+ = [E+i0^+ - V^2G_s^+]^{-1}\,.
\end{equation}
The solution of Eq. (\ref{Gbb}) is elementary and reads
\begin{equation}
 G^+_s=\frac {E}{2V^2}-\frac {i}{2V^2}\sqrt{4V^2-E^2}\,,
\end{equation}
for $E^2<4V^2$ and 
\begin{equation}
 G^+_s=\frac {E}{2V^2}-\frac {\sgn{E}}{2V^2}\sqrt{E^2-4V^2}\,,
\end{equation}
for $E^2>4V^2$.

Our goal is to study the STM current at finite bias, which is a particular case of
non-equilibrium transport. This is done using the non-equilibrium 
Green's function method or Keldysh method. This method is particularly suited to study the regime where
the system has a strong departure from equilibrium, such as when the bias potential
$V_b$ is large. We consider, however, that the system is in the steady state.  Since the seminal paper of Caroli
{\it et al.} on non-equilibrium quantum transport,\cite{Caroli} that the
method of non-equilibrium Green's functions started to be generalized to the
calculation of transport quantities of nanostructures.
There are many places where one can find
a description of the method,\cite{Ferry,Jauho} 
but a recent and elegant one was introduced
in the context of transport through systems that have bound states, showing that
the problem can be reduced to the solution of kind of quantum Langevin equation.\cite{Sen}

The general idea in this method is that two perfect leads are coupled to our system,
which is usually called the device. In our case the device is defined by the last
atom at the tip of the microscope. The Green's function of the
device has to be computed in the presence of the bulk of the tip and of graphene. 
This corresponds to our $G_{00}^+$ Green's
function. Besides the Green's function we need the effective coupling between the last atom
of the tip and the bulk atoms as well as that to the graphene atoms, 
which are determined in terms of the self-energies
\begin{equation}
\Gamma_{L/R} = \frac {i}{2\pi}(\Sigma^+_{L/R}-\Sigma^-_{L/R})\,. 
\end{equation}
Therefore the effective coupling $\Gamma_{L/R}$ depends on the surface Green's
function of the tip and of graphene. According to the general theory, the two systems
(bulk of the tip and graphene) are in thermal
equilibrium at temperatures $T_{L/R}$ and chemical potential $\mu_{L/R}$ and are connected
to the system at some time $t_0$. The bottom line is that the total current through the
device is given by (both spins included)
\begin{equation}
J=\frac {2e}{h}\int_{-\infty}^\infty dE T(E)[f(E,\mu_L,T_L)-f(E,\mu_R,T_R)]\,, 
\end{equation}
where $f(x)$ is the Fermi-Dirac distribution and
the transmission $T(E)$ is given by 
\begin{equation}
 T(E)= 4\pi^2\tr[\Gamma_LG^+_{00}\Gamma_RG^-_{00}]\,.
\end{equation}
Performing the trace (which in this case is only a product of complex numbers) we obtain
\begin{equation}
T(E)=4W_1^2W_2^2\Im G_s^+\Im G_{xx}^+\abs{G^+_{00}}^2\,. 
\end{equation}

\begin{figure}[htf]
\begin{center}
\includegraphics*[width=8cm]{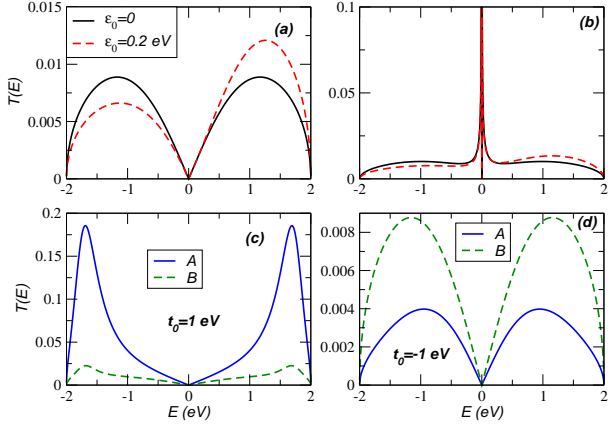}
\end{center}
\caption{(color on line)
Transmission coefficient $T(E)$ as function of the energy for zero
bias. The fixed parameters are:
$t=3.0$, $V=1$, $W_1=0.9$, $W_2=0.2$, all in electron-volt.
Panel (a) depicts the case of clean graphene and two different
values of $\epsilon_0$. Panel (b) depicts $T(E)$ through a $B$
site  when a vacancy sits at the $A$ site. Panel (c) depicts
$T(E)$ through $A$ and $B$ sites, when an impurity sits
at the $A$ site, taking $t_0=1$ eV and $\epsilon_0=0$. Panel (d) is the same
as panel (c) for $t_0=-1$ eV.
\label{fig:TE_all_cases}}
\end{figure}
Figure \ref{fig:TE_all_cases} represents $T(E)$ in several
conditions. In the clean case, we see the effect
the Dirac point has on the transmission, leading to a suppression
of the tunnelling for $E\simeq 0$. It is also clear that the parameters
characterizing the last atom of the tip strongly influences the form
of $T(E)$, leading to an asymmetry between negative and positive 
energies due to the finiteness of $\epsilon_0$. When a vacancy
is present at the $A$ site, the tunneling through the nearest $B$
sites is strongly modified relatively to the clean case, with
strong tunneling taking place at energies very close to the Dirac
point. When an impurity atom sits at the $A$ site the behavior
of the tunneling probability depends on the coupling between the
impurity and the carbon atoms. In the case of weak coupling to the
carbon atoms ($t_0>0$) the tunneling through the $A$ atom
is facilitated relative to the tunneling between the nearest 
carbon atoms. When the impurity is strongly coupled to the
carbon atoms ($t_0<0$), the reverse happens. This behavior
is easily understood, since strong (weak) coupling to the carbon atoms
is equivalent to a weak (strong) coupling to the tip of the microscope,
leading to weaker (stronger) tunnelling probability from the tip
to graphene.

\begin{figure}[htf]
\begin{center}
\includegraphics*[width=8cm]{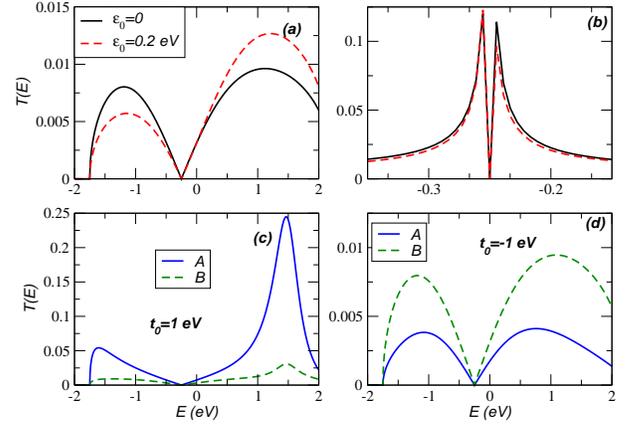}
\end{center}
\caption{(color on line)
Transmission coefficient $T(E)$ as function of the energy for finite
bias $V_b=0.5$ volt. The parameters and the description
are the same listed in the caption
of Fig. \ref{fig:TE_all_cases}. The energy scale in the case of the
vacancy was increased in order to see the low energy behavior
close to the Dirac point. 
\label{fig:TE_all_cases_V}}
\end{figure}

All the above discussion applies to zero bias voltage. When 
finite bias voltage is applied between the tip and graphene, the
behavior of the curves change. We will consider that the chemical
potentials of the tip and of graphene differ by the electrostatic energy
$eV_b$, where $V_b$ is bias potential. This amounts to change the
on-site energies relative to their value in equilibrium. We choose
to change the on-site energies of the tip by $eV_b/2$ and those
of the carbon atoms by $-eV_b/2$. In Figure \ref{fig:TE_all_cases_V}
we depict $T(E)$ for a finite value of the bias $V_b$. The most
distinctive difference relative to the case of zero bias
is the energy asymmetry induced by the bias relatively the
case of zero bias, even when $\epsilon_0=0$. 

\begin{figure}[htf]
\begin{center}
\includegraphics*[width=8cm]{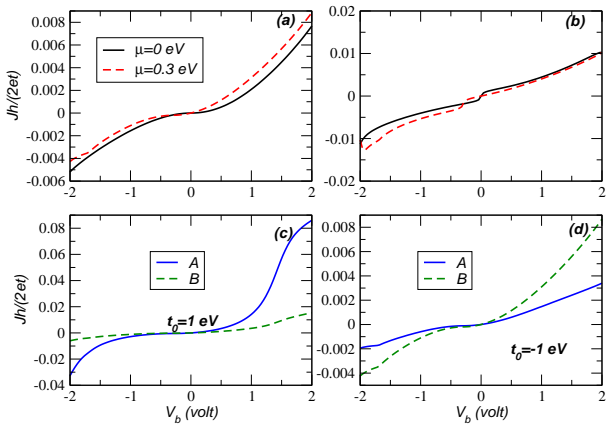}
\end{center}
\caption{(color on line)
Current $J$ as function of the bias voltage $V_b$. The parameters 
are the same listed in the caption
of Fig. \ref{fig:TE_all_cases} and $\epsilon_0=0.2$ eV. 
Panel (a) represents the clean the case for two values of the chemical potential $\mu$.
The case of the vacancy is represented in panel (b), also for two values of the chemical
potential. Panel (c) represents the case $t_0=1$ eV for $\mu=0.3$ eV, and panel
(d) is the same as (c) for  $t_0=-1$ eV. In panels (c) and (d), $A$ and $B$
refer to tunneling through the atom siting on sub-lattice $A$ or $B$, respectively.
\label{fig:TE_J}}
\end{figure}

The calculation of the current also depends on the chemical potential.
In the case of graphene this can be tuned by a back gate $V_g$. The relation
between the back gate voltage and the chemical potential of graphene
obeys the relation
\begin{equation}
 \mu = v_F\hbar\sqrt{\frac{\pi\epsilon\epsilon_0 V_g}{de}}\,,
\label{mu}
\end{equation}
which is numerically equal to $\mu=0.03\sqrt V_g$ in electron-volt.
The parameters in Eq. (\ref{mu}) are $\epsilon=3.9$, the dielectric constant
of SiO$_2$, $d=300$ nm, the thickness of the SiO$_2$ substrate, and $e$ the elementary
charge. Taking a typical value of $V_g=100$ V we obtain $\mu=0.3$ eV, which is the
value we assume for the chemical potential in the calculations below. Since we will
perform our calculations at zero temperature, the current is given by

\begin{equation}
J=\frac {2e}{h}\int_{\mu+eV_b/2}^{\mu-eV_b/2} dE\; T(E,V_b)\,. 
\end{equation}
The results for the calculation of $J$ versus $V_b$ is depicted in Fig. \ref{fig:TE_J}.
Looking at them, we see that there is an asymmetry between the
negative and positive values of $V_b$. For the clean case, it is clear
that the magnitude of the $J$ current does not depend much on the 
chemical potential, and therefore on the gate voltage $V_g$. The
same is true for the vacancy case. Comparing the curves for the
vacancy and for the clean case, we see that the order of magnitude
of the current is the same, but the shape of the curves is notoriously
different from that of the clean case. When the impurity is
weekly coupled ($t_0>0$) to the carbon atoms, the tunneling current is
facilitated through the impurity atom, as can be seen from
panel (c) of Fig. \ref{fig:TE_J}, with an absolute value five times larger
for $V_b$=1 volt. Also a clear jump is seen in $J$ (panel (c)), a
finger print of the resonances in the density of states seen in panel (c)
of Fig. \ref{fig:LDOS_func_R}. One note that in Fig. \ref{fig:LDOS_func_R} the
resonances are symmetrically positioned relatively to the Dirac point, but this is
not so for the jumps in the current. The reason is due to finite on-site energy
at the atom in the tip of the microscope.
On the other hand for strong coupling,
panel (d) of Fig. \ref{fig:TE_J}, there is not much difference from the clean case,
apart from the fact that tunneling current through the impurity or the
neighbor carbon atom has different magnitude.

\begin{figure}[htf]
\begin{center}
\includegraphics*[width=8cm]{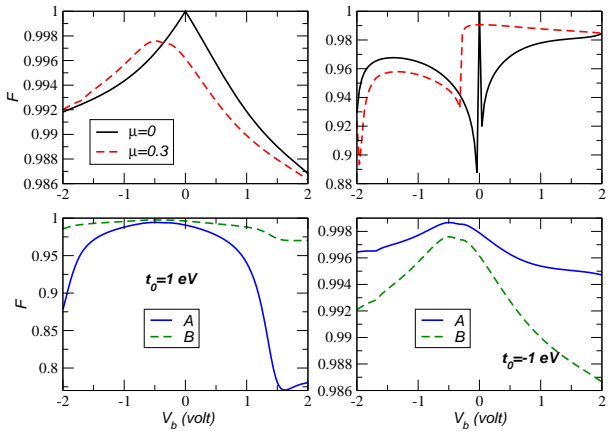}
\end{center}
\caption{(color on line)
Fano factor $F$ as function of the bias voltage $V_b$. The parameters 
are the same listed in the caption
of Fig. \ref{fig:TE_all_cases} and $\epsilon_0=0.2$ eV. 
Panel (a) represents the clean case for two values of the chemical potential $\mu=$0
and $\mu=$0.3 eV.
The case of the vacancy is represented in panel (b), also for two values of the chemical
potential. Panel (c) represents the case $t_0=1$ eV for $\mu=0.3$ eV, and panel
(d) is the same as (c) for  $t_0=-1$ eV. In panels (c) and (d), $A$ and $B$
refer to tunneling through the atom siting on sub-lattice $A$ or $B$, respectively.
\label{fig:Fano}}
\end{figure}

Another important quantity to fully characterize the STM current 
is the shot noise.\cite{Beenakker} For interacting systems this
quantity gives information on the possible existence of quasi-particles
with fractional charge. On disordered systems with no interactions,
information on transport open channels can be obtained. For fermionic
non-interacting electrons at zero temperature the
shot noise is defined as\cite{Buttiker}
\begin{equation}
 S=\frac{2e^2}{\hbar}\int_{\mu_R}^{\mu_L}dE\;T(E)[1-T(E)]\,.
\end{equation}
The relevant quantity is not $S$ directly but the Fano factor\cite{Beenakker}
defined as
\begin{equation}
 F=\frac {S}{eJ}\,.
\end{equation}
When the transmission $T(E)$ is strongly reduced we have $F\rightarrow 1$, and 
noise is said poissonian. On the other hand, if the system has a finite
density of open channels, $T(E)\rightarrow 1$, we have $F<1$ due to $[1-T(E)]\ll 1$. 
In Figure \ref{fig:Fano} we plot the Fano factor 
as function of the bias voltage.

Clearly, for all case but the vacancy, the Fano factor is close to
the poissonian values $F=1$. When $V_b\rightarrow 0$ and the chemical
potential is zero, the value of $T(E)$ quite small due to the presence
of the Dirac point, and the Fano factor take the limiting value $F=1$.
For the vacancy, we obtain much smaller values of $F$, which is indication
of the presence of the strong resonance seen  in the local density of states
for this case, which leads to an increase of the transmission.
It should be noted that in the cases of finite $t_0$, the concavity of the
curves for $F$ as function of $V_b$ allows to distinguish the case
of week coupling $t_0>0$ from the case of strong coupling $t_0<0$.

\section{Discussion and conclusions}
\label{conclusion}

We have studied in detail the electronic local density of states in graphene
close to a substituting atom. This quantity turns out to be important in
STM experiments, since the tunneling of the electrons from tip of the microscope
to the sample is determined by this quantity. The shape and magnitude
of the tunneling current depends on the nature of the substituting atom. In the
case of the vacancy the shape of the tunneling current has a very different signature
from the other cases. Also the magnitude of the current depends on 
the strength of the bonding between the impurity and the carbon atom.

The curves we have presented here are only indicative of the general behavior
of the density of states and of the tunneling current. In order to obtain
experimentally relevant curves we should have real numbers for the
parameters of the tip, with special importance for the last atom of tip.
Also the parameter $W_2$ could be modeled more accurately by introducing
the spatial dependence between the tip and the graphene surface. Density
functional calculations can be used to model the tip in realistic way.

In our analysis we have consider that the electrons can only tunnel
from the tip to the atom underneath, but when the tip is between two atoms
the tunnel will take place to more than one atom. However, in this case
the amplitude $W_2$ is strongly reduced and the tunneling current will
show a minimum. In fact it can be shown that the self energy has in this
case the form
\begin{equation}
\Sigma^+_R=\tilde W_2^2(G_{AA}^++G_{BB}^+)\,, 
\end{equation}
assuming the tip exactly in between the $A$ and $B$ atoms, and therefore
the value of the current will be controlled mainly by the value of 
$\tilde W_2$.

As we have seen in the calculation of the density of states, the case
of weak coupling leads to the appearance of resonances at large
energy values. The scan of the bias in our calculations did not reach these
energies, but if it does a clear signature will be seen in the tunneling current.
Also, for fixed bias, moving the tip away form the impurity will show oscillations
in the STM current.

One ingredient not included in our model is
the effect of on-site energies at the impurities.\cite{charlier2} 
As a consequence, a natural question
is whether our results are strongly modified if this effect is included.
One should note that the values of the on-site
energy and the hopping are correlated with each other in the case
of boron and nitrogen impurities. When the hopping in enhanced the on-site
energy is positive relatively to that of the carbon atoms. In this case,
the results of our calculations in the present work show no
qualitative changes from the more general model, as more detailed calculations show.
In the case of reduced hopping, the same resonances we see in the local density
of states of our work are also present in the more general approach, but
their position in energy changes and an asymmetry of those resonances relatively to
the Dirac point develops. One should also note that the different chemical nature of the atoms
making the tip of the microscope also induces an asymmetry in the current, even when
the LDOS does not show such asymmetry. Therefore, a careful study is needed to disentangle
the effects from the tip and from a finite on-site energy at the impurity
atoms.

Another relevant question is whether the the resonances see in the
density of state will broaden so much when the concentration of impurities
increases leaving no trace of them in the LDOS.
Our calculations naturally refer to the diluted limit, where the distance
between impurities is large. When the concentration increases the resonances
will broaden, but their finger prints still remains in the LDOS.\cite{vitor} 
In Ref. \onlinecite{vitor},  Fig. 13(a) shows that
the resonances remain well defined even for concentration of impurities up to
10\%. It is not possible  to give a characteristic length at the Dirac point such that
above it the impurities could be considered to act isolated. This is so because Fermi momentum
is zero. One was therefore to rely on numerical calculations with a varying concentration
of impurities as in Ref. \onlinecite{vitor}.

\section*{Acknowledgements}
This work was supported by FCT under the grant PTDC/FIS/64404/2006.
J. M. B. Lopes dos Santos is acknowledged for fruitful discussions.

\appendix

\section{Free propagators and useful relations}
\label{free}

The propagators for the perfect lattice are ($\hbar=1$)
\begin{eqnarray}
G_{AA}^0(\omega_n,\bm k,\bm p) = \frac {i\omega_n\delta_{\bm k,\bm p}}
{(i\omega_n)^2-t^2\vert\phi(\bm k)\vert^2 }\,,\\
G_{BA}^0(\omega_n,\bm k,\bm p) = \frac {\delta_{\bm k,\bm p} \phi^\ast (\bm k)}
{(i\omega_n)^2-t^2\vert\phi(\bm k)\vert^2 }\,,\\
G_{BB}^0(\omega_n,\bm k,\bm p) = \frac {i\omega_n\delta_{\bm k,\bm p}}
{(i\omega_n)^2-t^2\vert\phi(\bm k)\vert^2 }\,,\\
G_{AB}^0(\omega_n,\bm k,\bm p) = \frac {\delta_{\bm k,\bm p} \phi (\bm k)}
{(i\omega_n)^2-t^2\vert\phi(\bm k)\vert^2 }\,.\\
\end{eqnarray}
From these, the integrals  (\ref{bargaa0}) and (\ref{tildegaa0}) are
defined as

\begin{equation}
\bar G^0_{AA}(\omega_n) = i\omega_n\int 
\frac {\rho(E)dE}
{(i\omega_n)^2-E^2}\,,
\end{equation}

\begin{eqnarray}
\tilde G^0_{AA}(\omega_n)&=&
 \frac  {i\omega_n}{t^2}\int 
\frac { E^2\rho(E)dE}
{(i\omega_n)^2-E^2}\nonumber\\
&=&-i\omega_n t^{-2} + (i\omega_n)^2t^{-2}\bar G^0_{AA}(\omega_n)
 \,,
\end{eqnarray}
and 
\begin{equation}
\rho(E) = \frac {A_c}{4\pi^2}\int_{BZ} d^2\bm k
\delta (E-t\vert \phi(\bm k)\vert)\,.
\label{dosE}
\end{equation}
where $A_c=3\sqrt 3 a_0/2$ is the area of the unit cell.
With the above definition $\rho(E)$ is finite in the energy range
$0<E<3t$.
In addition we also have the useful relation
\begin{eqnarray}
\sum_{\bm k}\phi(\bm k)G^0_{AA}(\omega_n,\bm k )
&=&\sum_{\bm k}\phi^\ast(\bm k)G^0_{AA}(\omega_n,\bm k )\nonumber\\
&=&\frac {N_c} 3 \tilde  G^0_{AA}(\omega_n) \,.
\label{niceproperty}
\end{eqnarray}
Similar equations hold for $G^0_{BB}$.


\end{document}